%% file: vpn_recom.tex
\documentclass[conference]{IEEEtran}
\IEEEoverridecommandlockouts
\usepackage{cite}
\usepackage{amsmath,amssymb,amsfonts}
\usepackage{algorithmic}
\usepackage{graphicx}
\usepackage{textcomp}
\usepackage{xcolor}
\usepackage{tikz}
\usepackage{pgfplots}
\usepackage{todonotes}
\usepackage{tabularx}

\def\BibTeX{{\rm B\kern-.05em{\sc i\kern-.025em b}\kern-.08em
    T\kern-.1667em\lower.7ex\hbox{E}\kern-.125emX}}
\tikzset{every picture/.style={line width=0.75pt}}
\tikzset{every picture/.style={line width=0.75pt}}
\tikzset{cross/.style={cross out, draw=black, fill=none, minimum size=2*(#1-\pgflinewidth), inner sep=0pt, outer sep=0pt}, cross/.default={2pt}}
\usetikzlibrary{positioning, shapes, trees, automata, calc, fit, automata, shapes.misc, arrows, patterns}
\begin{document}
\title{Recommendations on using VPN over SATCOM}

\author{
\IEEEauthorblockN{Romain Guilloteau}
\IEEEauthorblockA{\textit{VIVERIS TECHNOLOGIES}}
\and
\IEEEauthorblockN{David Pradas}
\IEEEauthorblockA{\textit{VIVERIS TECHNOLOGIES}}
\and
\IEEEauthorblockN{Guillaume Pelat}
\IEEEauthorblockA{\textit{VIVERIS TECHNOLOGIES}}
\and
\IEEEauthorblockN{Nicolas Kuhn}
\IEEEauthorblockA{\textit{CNES}}
}

\maketitle

\begin{abstract}

VPN are a secured tunnel that help service providers to exchange data over
	non-secured networks. There is a large variety of VPN solutions
	that have variable deployment impacts on the target architecture as well 
	as performance limitations or opportunities. 

This technical report compares Wireguard and OpenVPN for various SATCOM
	deployment scenarios and topologies. 

\end{abstract}


\input{introduction}

\input{scenario}
\input{results_single_flow}

\input{results_multi_flow}

\bibliographystyle{IEEEtran}
\bibliography{reference}

\end{document}

%% file: introduction.tex
\section{Introduction}
\label{sec:introduction}

Working from home or interconnecting entreprise networks increase security
needs, and this can be fulfilled by taking advantage of VPN solutions. When using
HTTPS is not enough, an added layer of security may be deployed to guarantee
that crossing a non-secured network will be safe.  

Since VPN can operate at the application or network layer, comparing VPN
solutions may not be straightforward. Wireguard~\cite{wireguard} operates at
the network layer and aims at replacing IPsec for most use cases.
OpenVPN~\cite{openvpn} works at the application layer and is not compatible with
IPsec. 

Depending on the scenarios of deployment, it may not be easy for an IT system
provider to guarantee end-to-end performances for one solution or the other one. 
As an example, Wireguard operating at the kernel level depends on
the machine on which it is deployed, while OpenVPN can be tuned to the deployment use
case by, e.g. using TCP or UDP to carry the secured tunnel traffic. 

SATCOM systems exhibit a wide variety, both in terms of delay and goodput,
making it hard to assess the relevant VPN solution for a specific
scenario~\cite{I-D.jones-tsvwg-transport-for-satellite}. Indeed, for the sake
of good end-to-end performance, SATCOM systems deploy Performance Enhancing
Proxies~\cite{RFC3135}. Exploiting VPN tunnels may result in by-passing these
proxies and impact the goodput of data transfers.

This technical report contributes to the performance evaluation of VPN tunnels.
We do not claim to provide the most extensive study on the subject, since we
focus on the SATCOM scenarios. That being said, the performance evaluation may
be of interest for other scenarios where performance enhancing proxies are
deployed.

%% file: scenario.tex
\section{Scenarios}
\label{sec:scenario}

This section describes the scenario that has been considered throughout this
evaluation study. It also presents the SATCOM system and the characteristics
of the VPN solutions.

\subsection{Architecture}
\label{subsec:archi}

\begin{figure*}[h]
        \centering
        \includegraphics[width =\linewidth]{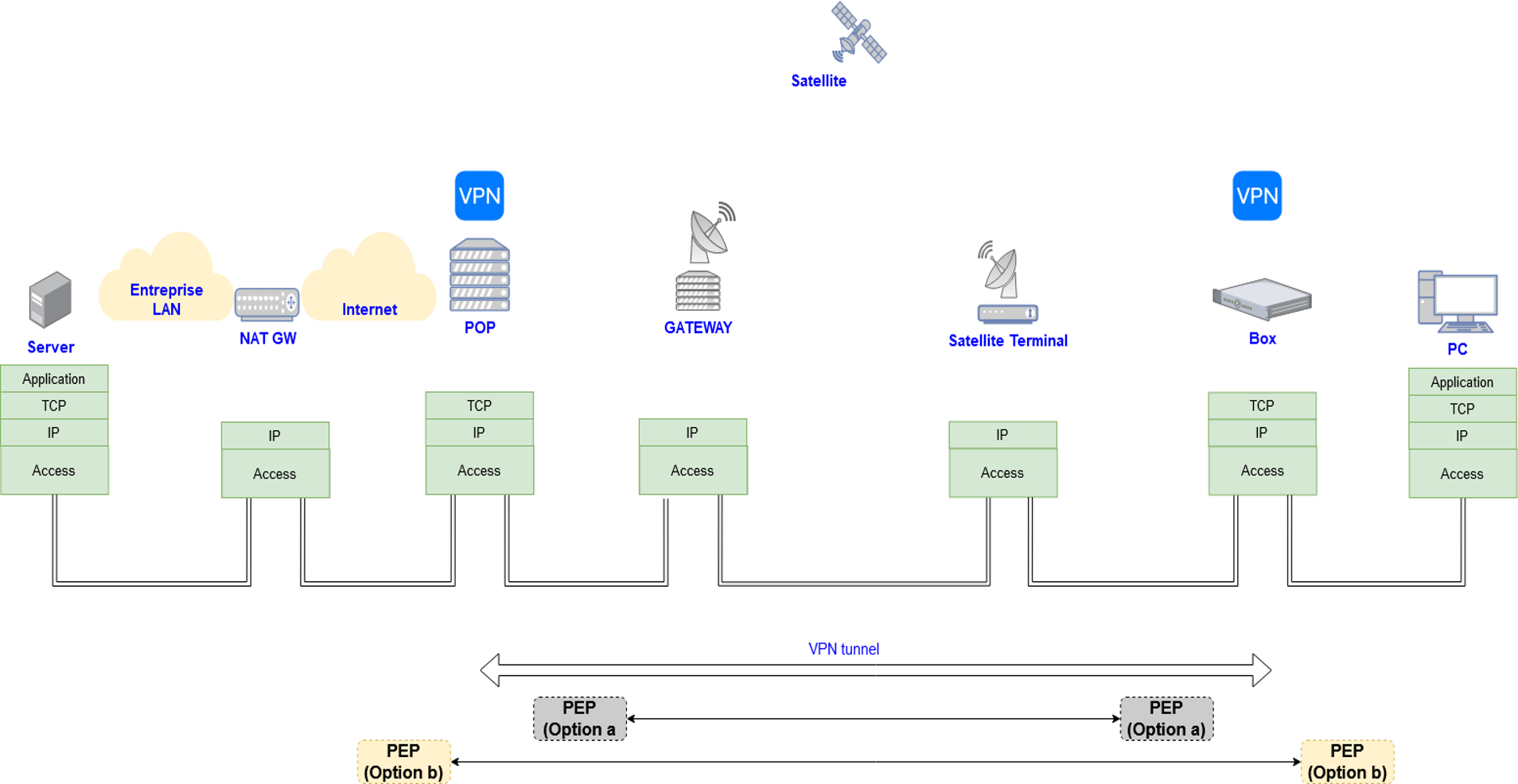}
        \caption{Platform architecture}
        \label{fig:platform-archi}
\end{figure*}

The architecture that will be exploited in this technical report is reported in
Figure~\ref{fig:platform-archi}.

An end user (PC) is connected to a box (offering Internet/media services 
and satellite access management) that includes the VPN client. A
satellite terminal, a satellite and a satellite gateway interconnect the
client's box to a POP (Point of Presence) that includes the VPN server. Then, the POP is connected
to the entreprise LAN that contains the server that the end user aims to reach.

The SATCOM solution might integrate a performance enhancing proxy that may be
either within or out of the VPN tunnel.

\subsection{SATCOM system}
\label{subsec:satcom}

The satellite system is emulated with \texttt{netem} and the tests are orchestrated with OpenBACH~\cite{openbach}. 
This technical report consider GEO and LEO systems. The characteristics of each system are the following:
\begin{itemize}
	\item GEO : RTT of $500$\,ms, bottleneck bandwidth of $10$\,Mbps 
	\item LEO : variable RTT, botteleneck bandwidth of $10$\,Mbps
\end{itemize}

End-to-end losses on SATCOM systems have been measured in~\cite{9268814}. In order to assess the impact of losses on the proposed solutions, we have also included random losses on the SATCOM system.

The PEP has been configured with the following options:
\begin{itemize}
	\item Before or after the VPN tunnel
	\item CUBIC, CUBIC without Hystart and BBRv2
	\item Various initial congestion windows
\end{itemize}

\subsection{VPN solutions}
\label{subsec:vpn}

The VPN solutions are either Wireguard or OpenVPN. In particular, the following parameters have been configured for OpenVPN:
\begin{itemize}
	\item Adapted socket buffer sizes (to match the Bandwidth-Delay-Product of the network)
	\item MTU/MSS size
	\item UDP or TCP mode
\end{itemize}

\subsection{Application layer}
\label{subsec:appli}

The following end-to-end applications have been introduced: 
\begin{itemize}
	\item File transfers with iperf3 
	\item Web transfers
	\item VoIP
\end{itemize}

%% file: results_single_flow.tex
\section{Single flow scenarios}
\label{sec:results_single_flow}

This section includes the results that have been obtained. A subset of the obtained results is presented and more results, based on the parameters listed in Section~\ref{sec:scenario}, are available upon request.

This section presents the results for a $30$\,MB file transfer when only one flow is considered. The options are the following: 
\begin{itemize}
    \item VPN : no VPN, OpenVPN (UDP and TCP) and Wireguard;
    \item PEP : no PEP, PEP in B position or PEP in A position; 
    \item Transport (applied end-to-end and on the PEP): CUBIC (with or without Hystart) or BBRv2.
\end{itemize}

\begin{figure*}[h]
        \centering
        \includegraphics[width =\linewidth]{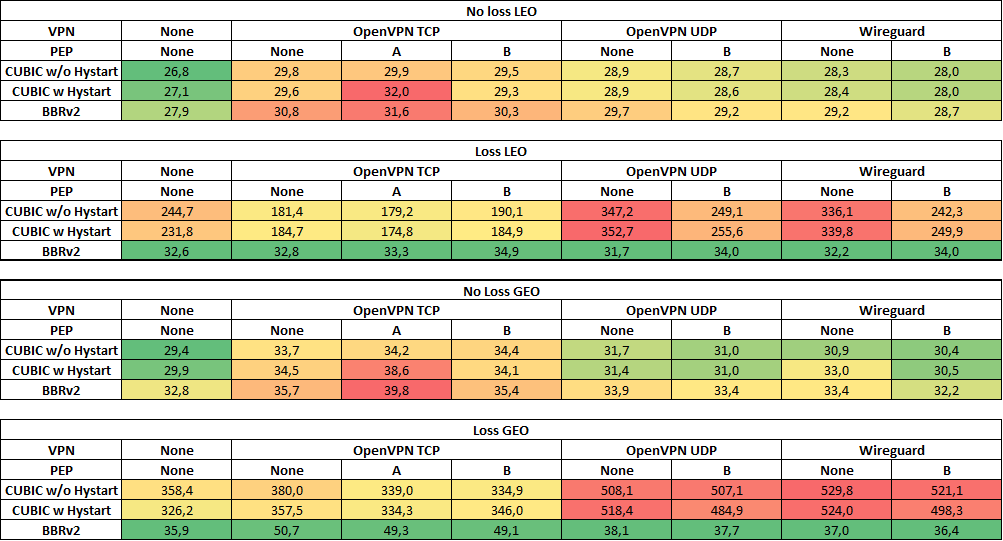}
	\caption{Summary of the results - $30$\,MB file transfer time (s)}
        \label{fig:base_test}
\end{figure*}

The results are shown in Figure~\ref{fig:base_test}. The table gathers all the
results. Comparative tables are shown in Figure~\ref{fig:no-loss-leo},
Figure~\ref{fig:no-loss-geo}, Figure~\ref{fig:loss-leo} and
Figure~\ref{fig:loss-geo}.

\subsection{No loss scenarios}

\begin{figure*}[h]
        \centering
        \includegraphics[width =\linewidth]{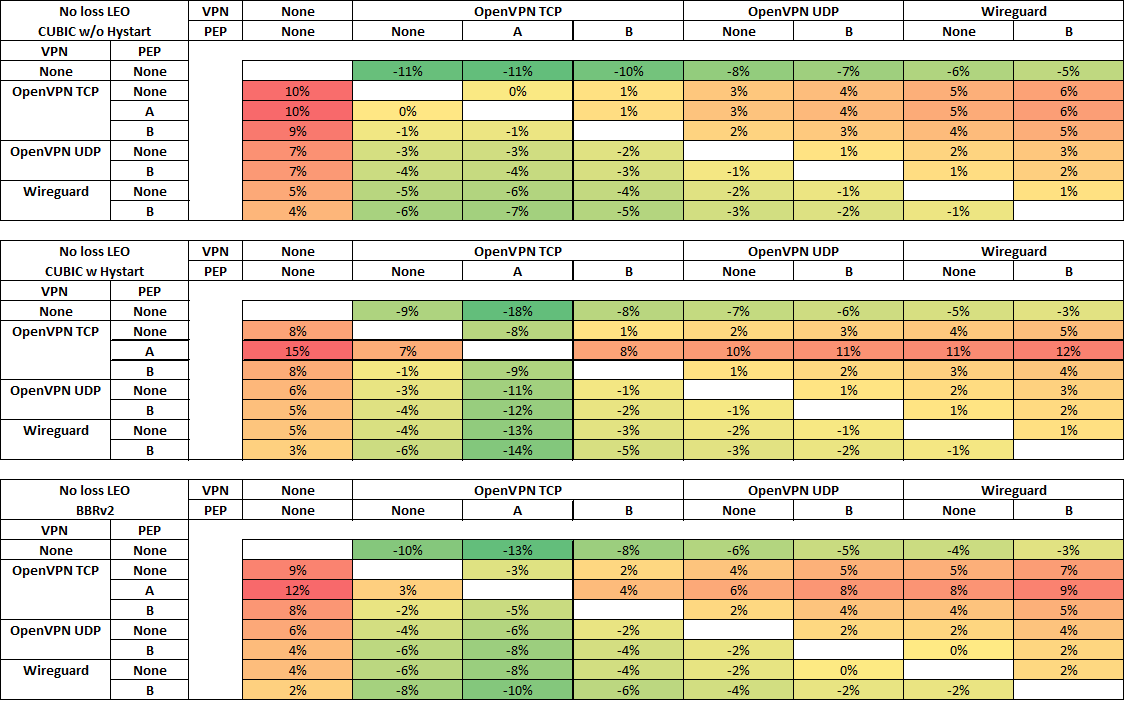}
        \caption{Comparative results - No loss LEO scenario}
        \label{fig:no-loss-leo}
\end{figure*}

\begin{figure*}[h]
        \centering
        \includegraphics[width =\linewidth]{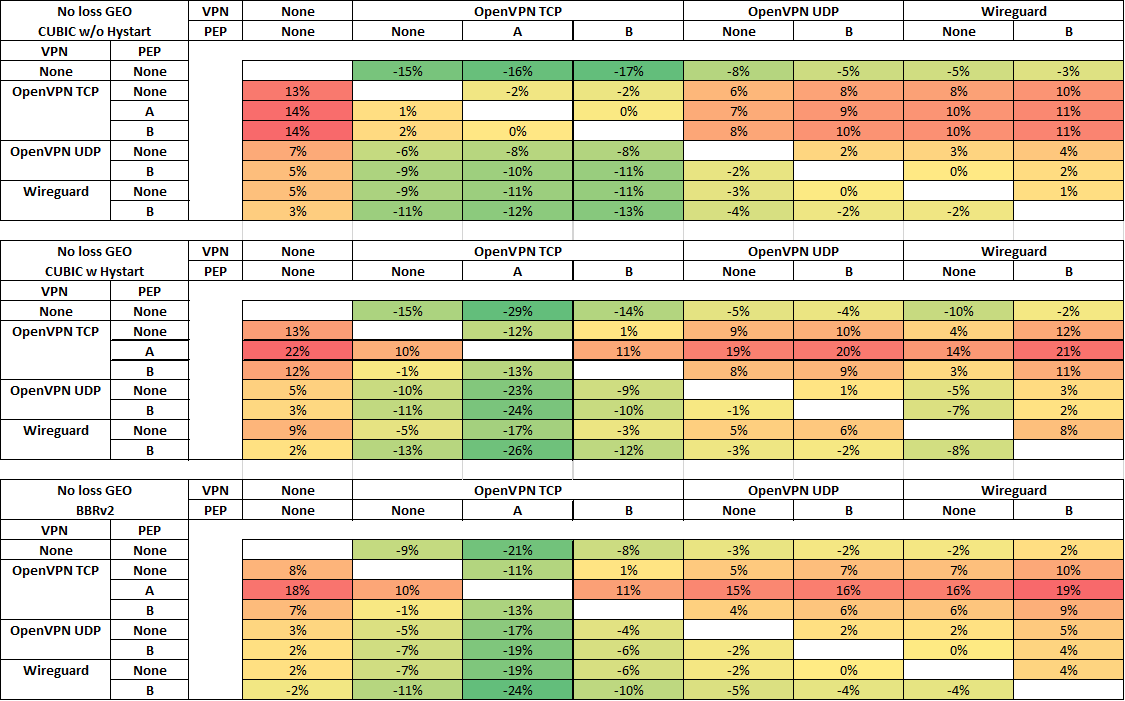}
        \caption{Comparative results - No loss GEO scenario}
        \label{fig:no-loss-geo}
\end{figure*}

The results are shown in Figure~\ref{fig:base_test}, 
Figure~\ref{fig:no-loss-leo} and Figure~\ref{fig:no-loss-geo}.

All transport layer variants are affected by the presence of a VPN, but the
difference can be neglicted. 

When a VPN has to be included, the solution based on OpenVPN TCP in PEP
position A results in the worst performance (i.e. "TCP in TCP" issue) by
increasing the transer time by at least $10$\,\%. 

When a VPN has to be included, it is then recommended to use Wireguard with a
PEP in position B. It is worth mentionning that other configurations, such as
OpenVPN UDP with/without PEP and Wireguard without PEP, exhibit fair
performance. Indeed, when we do not consider the OpenVPN TCP in PEP position A
configuration, the time to transfer is further reduced by only up to $5$\,\%
with Wireguard with PEP in position B if compared with the one achieved with
aforementionned configurations. 

\subsection{Loss scenarios}

\begin{figure*}[h]
        \centering
        \includegraphics[width =\linewidth]{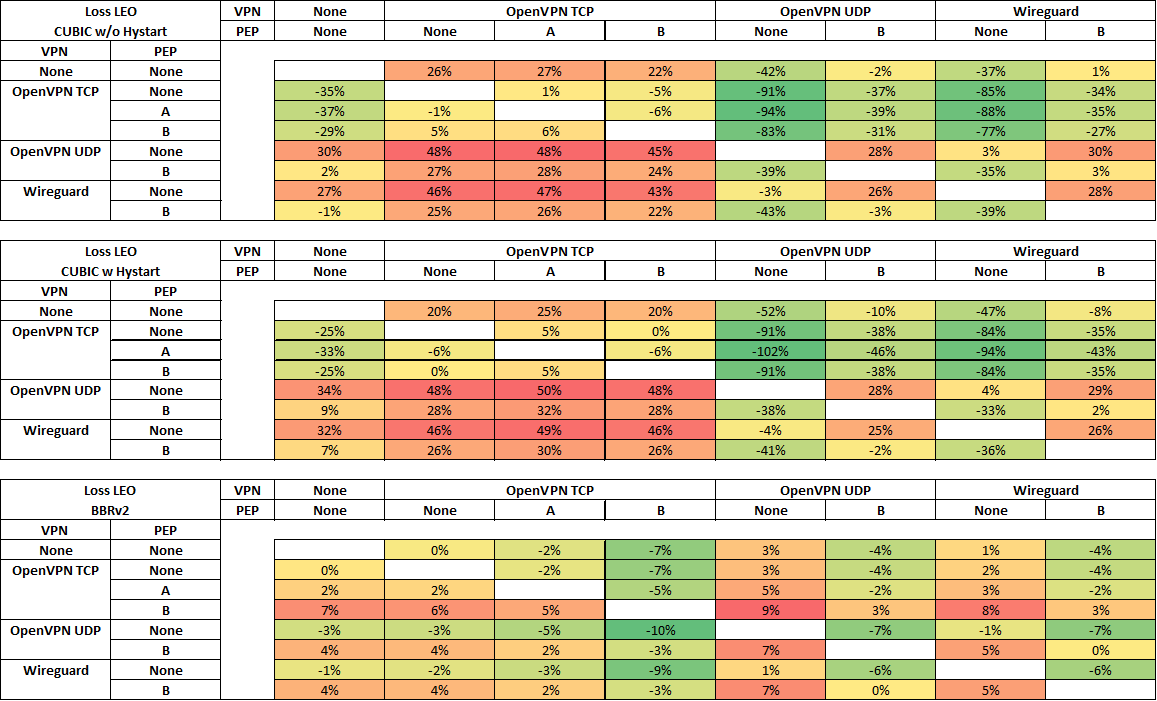}
        \caption{Comparative results - Loss LEO scenario}
        \label{fig:loss-leo}
\end{figure*}

\begin{figure*}[h]
        \centering
        \includegraphics[width =\linewidth]{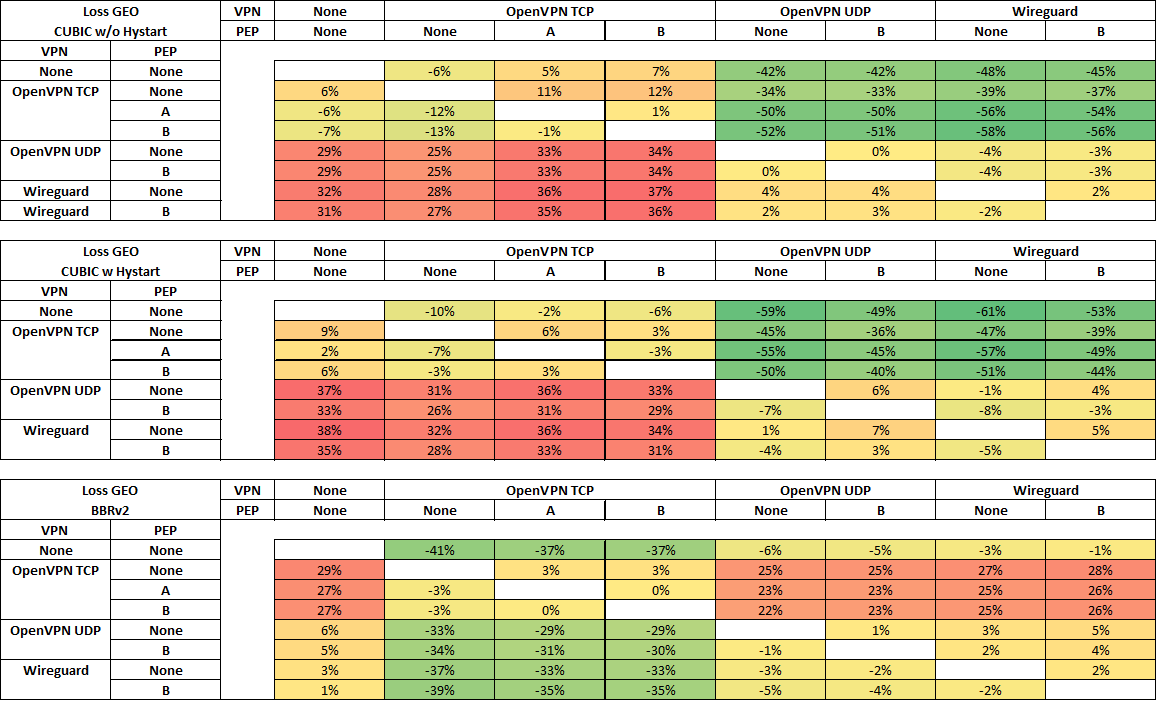}
        \caption{Comparative results - Loss GEO scenario}
        \label{fig:loss-geo}
\end{figure*}

The results are shown in Figure~\ref{fig:base_test},
Figure~\ref{fig:loss-leo} and Figure~\ref{fig:loss-geo}.

When there are losses on the satellite link, the choice of BBRv2 as a transport
layer protocol helps in improving well-known TCP CUBIC issues when lossy links
are considered. In this case, OpenVPN UDP or Wireguard show the best performances 
specially for GEO satellites. However, there are cases where the end-to-end congestion
control can not be adapted (e.g. with end-to-end QUIC flows or end servers
non-managed by the operator).

In the case where the end-to-end transport is CUBIC, OpenVPN TCP exhibits the
best performance by reducing the transfer time by $30$\% as opposed to the case
where no VPN and no PEP is proposed for the LEO scenario. In the case where the
end-to-end transport is CUBIC, for the GEO scenario, OpenVPN TCP also exhibits
the best configuration.   

The recommendations that are proposed in this section are mainly related to the
position of the losses. They are currently emulated in the satellite link and
other position (e.g. losses on the LAN link) may differ. In particular, the
gains brought by PEP when losses are located at the last-mile would be more
important.

%% file: results_multi_flow.tex
\section{Multiple flow scenarios}

This document reported so far results on the single flow scenarios and on the
scenarios where the congestion controls where the same on all entities. In the
framework of this study, experiments with multiple flows and with congestion
controls that are different on the end host and on the PEP have been also done. This
section sums up the main conclusions of this extended activity and the details
may be provided up-on request.

\subsection{CUBIC on the end hosts and BBRv2 on the PEP}

In order to have more confidence on the recommendation on exploiting BBRv2
whenever it can be applied, it can be noted that the experiments considered
BBRv2 on the end hosts and on the PEP. As a result, we can hardly guarantee
that the conclusions would be the same if BBRv2 was applied only on the PEP,
while CUBIC is applied on the end host. We have compared the downloading time
of a 30 MB file in the case where BBRv2 is applied on all entities and in the
case where BBRv2 is applied everywhere but the end host. 

The only case where the performance are not acceptable is the one where CUBIC
is applied end-to-end and the PEP could not split the traffic (and apply BBRv2
on the lossy segment of the network). When the connection can be split, the
gains brought by BBRv2 tolerance to packet loss are present: the transfer time
oscillates between 32 and 46 seconds.

As a result, when BBRv2 can be exploited on the lossy segments, using it
whenever possible remains the recommendation of this paper.

\subsection{Impact of using multiple flows}

To assess the impact of using multiple flows across the network, we have
considered:
\begin{itemize}
	\item for the loss-less scenarios: a flow that transmits data during
		hundred seconds and a second flow starting ten seconds later
		that transmits data during eighty seconds. 
	\item for the loss scenarios: a flow that start an unlimited data
		transfer and a second flow transmit thirty MB ten seconds
		later. 
\end{itemize}

In the loss-less scenarios, Wireguard with a PEP in a B position enabled the
largest amount of data transmitted in general and did not increase considerably
the unfairness between the two flows. In all configurations of PEP and VPN, the
capacity sharing between the flows oscillates between 40 \% and 60 \%. 

In the loss scenarios, Wireguard also exhibited the best performance in terms
of 30 MB transfer time and did not impact the existing unfairness between the
two flows that were considered. 

OpenVPN TCP did not exhibit good performance in the multiple flow scenarios,
which may resides in the fact that multiple TCP flows are tunneled within one
single TCP flow. As a result, when using OpenVPN TCP, it is recommended to
consider one tunnel per TCP flow.